# Spatial-temporal water area monitoring of Miyun Reservoir using remote sensing imagery from 1984 to 2020

Chang Liu[1, 2, 3], Hairong Tang[1, 2, 3], Luyan Ji[1, 3*], Yongchao Zhao[1, 2, 3]

1. *Aerospace Information Research Institute, Chinese Academy of Sciences, Beijing 100094, China;*
2. *University of Chinese Academy of Sciences, Beijing 100049, China;*
3. *Key Laboratory of Technology in Geo-Spatial Information Processing and Application System, Beijing 100190, China;*
* *Correspondence: jily@mail.ustc.edu.cn*

**Abstract:** Miyun Reservoir has produced huge benefits in flood control, agricultural irrigation, power generation, aquaculture, tourism, and urban water supply. Accurately water mapping is of great significance to the ecological environment monitoring of the Miyun Reservoir and the management of the South-to-North Water Diversion Project. On the 60th anniversary of the completion of the Miyun Reservoir, we took the Miyun Reservoir as the study area and collected all the Landsat-5 and Landsat-8 remote sensing images from 1984 to 2020 for water mapping. Based on the spectral, topographical and temporal-spatial characteristics of water, we proposed an automated method for long-term researvoir mapping, which can solve the problems caused by cloud, shadow, ice and snow pixels. Moreover, it can also deal with the "same objects with different spectra" and spectral mixed problems. The overall accuracy is as high as 98.2% for the case with no cloud or snow/ice cover. The landscape division index is introduced to analyze the morphological changes of Miyun Reservoir. Based on the mapping results, it is suggested that the water area change of the reservoir can be divided into five periods: "growth period" (1984-1993), "peak period" (1994-1999), "decline period" (2000-2003), "protection period" "(2004-2014), and "recovery period" (2015-2020), during which the east and west areas of the reservoir have experienced the process of "division-combination-division-combination". The largest reservoir area reached 151.6 km² (1995), and the smallest area was only 57.3 km² (2004). Moreover, we also find that the largest area in a year occurs from August to September, and the smallest area occurs in May. In addition, we have found that the changes of water area of Miyun Reservoir are closely connected to urban development, government policies and precipitation varitations. The study shows that our method used here are potentially extendable to other reservoirs for long-term area monitoring.

area, landscape index, Landsat

**Supported by** The second comprehensive scientific investigation of the Qinghai-Tibet Plateau (2019QZKK0206); National Natural Science Foundation of China (No. 61701477); National Natural Science Foundation of China Youth Fund(NO.61805246)

# 1 Introduction

Beijing is not only the capital, but also the political and cultural center of China. However, its available freshwater resources are extremely scarce, and the supply pressure of water resources is huge. Miyun Reservoir is the largest source of drinking water in Beijing (Li et al, 2004) and supplies at least tens of millions of people. In addition, the Reservoir also undertakes the industrial water supply responsibilities of many factories and enterprises, including the Sinopec yanshan petrochemical company, Shougang group, etc. Great benefits have been generated in many aspects after the completion of Miyun Reservoir, including flood control, agricultural irrigation, power generation, aquaculture, tourism, urban water supply, etc (Hu et al, 2007). Therefore, it is of great significance for monitoring and managing the ecological environment of Miyun Reservoir to accurately complete its mapping and analyze its characteristics and spatial-temporal changes.

Due to the late start and immaturity of reservoir ecological environment protection and the lack of scientific monitoring and treatment methods in China, the reservoirs can not give full play to its role (Zhang, 2019). Reservoir management has a low degree of integration with information technology and the monitoring and management such as mapping is still mainly completed manually, which inhibits the effectiveness of reservoir ecological environment protection and management in China (Li, 2019). Benefit from the development of optical remote sensing technology and the accumulation of high-quality data, long-term dynamic analysis with remote sensing data has become an important means of reservoir monitoring.

At present, the remote sensing research on Miyun Reservoir mainly focuses on the water quality of the reservoir, the landscape ecology of the upstream basin and the vegetation analysis of the reservoir shore zone (Hu et al, 2019; Zhang et al, 2020; Ma et al, 2020). However, the research on reservoir water surface information monitoring is very lacking, only Cao Ronglong et al(2008）adopt water index method with seven Landsat images to map Miyun Reservoir every four years from 1984 to 2005. But the mapping results are old, the mapping frequency is low, the extraction method is simple and the accuracy is not very high. What's more, with the water supply of the Middle Route Project of the South-to-North Water Diversion at the end of 2014, the water surface information of Miyun Reservoir presents new characteristics, which requires more scientific monitoring and management.(Li et al, 2019).

Water extraction using remote sensing data mainly uses the spectral characteristics of water, that is, the reflectivity difference between water and vegetation, soil and other ground objects. The existing methods of reservoir water information extraction can be classified into three categories: single-band threshold method, multi-band method (including spectral relationship and water index), and extraction methods after classification (Liu et al, 2013). The single-band threshold method uses the short wave infrared band threshold segmentation of remote sensing image to extract the reservoir water surface, which is difficult to eliminate the influence of many ground objects and has low accuracy. The extraction methods after classification include supervised classification, unsupervised classification and so on. Most of them focus on the extraction of water from single scene images and rely heavily on manual samples. At present, the most widely used methods are the multi-band ones, in which the spectral relationship method enhances the water information by constructing the dominant combination of different bands and separates the water from the remote sensing image. However, the model and threshold selection of the spectral relationship method are greatly affected by the data and ground object spectra, and cloud and water are commonly confused. The water index method enhances the contrast between water and other ground objects by arithmetic operation on the wave band. It carries out water mapping combined with threshold segmentation. The water index method is a relatively fast and reliable means for extracting long-time water information(Liu et al, 2016). The commonly used water indexes include Normalized Difference Water Index (S. K. McFEETERS et al., 1996), Modified Normalized Difference Water Index (Xu et al., 2006), Comprehensive index of normalized vegetation and water (Lu et al., 2011), Automatic Water Extraction Index (Feyisa et al., 2014) and so on. The existing studies based on the above indexes mainly focus on single phase images, but there are few studies on reservoir water surface information monitoring using long-time series remote sensing images.

The existing water mapping products covering Miyun Reservoir area mainly include as follows: 1) China's Lake products (Ma Ronglong et al., 2011) from 1960s to 2008. There are only 1-2 scenes per year in this product and it can not effectively capture the dynamic characteristics of Miyun reservoir. The products rely on manual interpretation and cannot be updated quickly; 2) Global water monthly mapping product with 30m resolution from 1984 to 2015 (pekel et al, 2016), which does not deal with clouds, ice, snow and mixed pixels; 3) The global 30m resolution inland water data set in 2000 (Feng et al, 2016), which has only one issue in 2000; 4) The improved water layer of the Finer resolution observation and monitoring of the global land cover (FROM-GLC) (Ji et al., 2015). The original FROM-GLC (Gong et al., 2013) uses Landsat images with supervised classification method (Support Vector Machine) to classify the surface into 10 categories (the sixth category is water body). The improved water maps reduced the impact of cloud and mountain shadow, but the mapping product has only one issue which in 2010; 5) Global water body daily mapping product from 2000 to 2016 (Ji et al, 2018), which is produced based on 500m resolution MODIS data. Compared with Miyun Reservoir in the study area, the spatial resolution of the product is not enough, the accuracy is difficult to meet the demand, and the time span is short. In addition, the above water products are large-scale mapping and the extraction accuracy in Miyun Reservoir area is difficult to be fully guaranteed.

Therefore, based on the Landsat reflectance data from 1984 to 2020, this paper solves the difficult problems such as cloud and snow interference, terrain shadow and mixed pixels in the reservoir dynamic mapping. We put forward a set of automatic and high-precision reservoir mapping method, which realizes the long-time, high-frequency and accurate mapping of Miyun Reservoir. Based on the historical mapping results of Miyun reservoir, we analyze the changes of water surface information such as area, coverage and landscape of Miyun reservoir from 1984 to 2020. In the last, we analyze the driving factors of the changes which can be referenced for reservoir monitoring and management.

# 2 Research area and data

## 2.1 Research area

Miyun Reservoir (40°23′N-40°34′N, 116°49′E-117° 4′E) is located in the Yanshan hills which are 13 kilometers north of Miyun County in the northeast of Beijing City. Miyun reservoir is the largest drinking water supply source in Beijing and the largest artificial lake in Asia, which is known as the *Pearl of Yanshan*. The reservoir was built in September 1960 with an area of about 180 square kilometers, a storage capacity of 4.375 billion cubic meters and an average depth of 30 meters. It is divided into two parts, East and West. Miyun reservoir is an important regulating reservoir with strong flood storage capacity by taking the water from the Chaobai River. Bai River originates from Guyuan County and is put into storage through Chicheng County, Yanqing County and Huairou District. Chao River originates from Fengning County,and is put into storage through Luanping county and ancient Beikou. The precipitation in the basin is rich and has obvious seasonal changes. The precipitation from July to August accounts for 70% of the whole year, mostly heavy rain and rainstorm(Li et al, 2004).

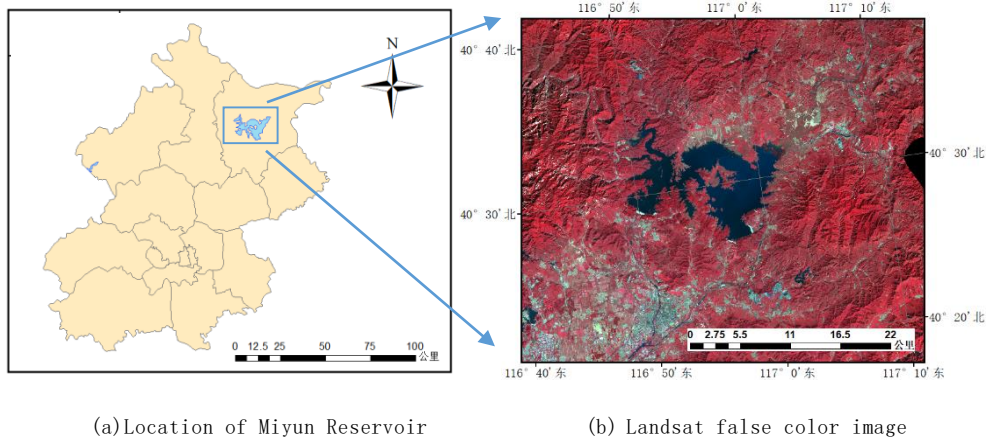

(a) Location of Miyun Reservoir  (b) Landsat false color image

Fig.1 The location of Miyun Reservoir in Beijing ((a) the boundary of the lakes from the Global Lakes and Wetlands Database (GLWD) (Lehner B. et al, 2004); (b) The false color image below is the same: R: Near-infrared, G: Red, B: Green; date: 2013.10.03)

**2.2 Data**

**2.2.1** Multi-spectral reflectance data

Landsat satellites have been observing the earth for 40 years and have played an important role in the monitoring and management of global water resources, food, forests and other natural resources (Jiang et al, 2013). At present, the satellite in service is Landsat-8 launched in 2013, while Landsat-5 was retired in June 2013 and Landsat-7 was launched in 1999 with equipment failure occurred in 2003. This study collected all Landsat-5 and Landsat-8 images (path = 123, row = 032) in Miyun Reservoir area from 1984 to 2020, including 485 Landsat-5 images and 166 landsat-8 images, as a total of 651. Figure 2 shows the annual and monthly cloud coverage of Landsat data in Miyun Reservoir area. It can be seen that the median cloud coverage of 10-year data exceeded 50% from 1984 to 2020. It can also be found from Figure 2 (b) that the problem of cloud is particularly serious in summer. As there is less effective ground information on images with high cloud coverage, this paper screen out Landsat images with cloud coverage of more than 80%. In addition, the images with geometric calibration errors and strip damage are removed. Finally, 474 scenes are actually used.。

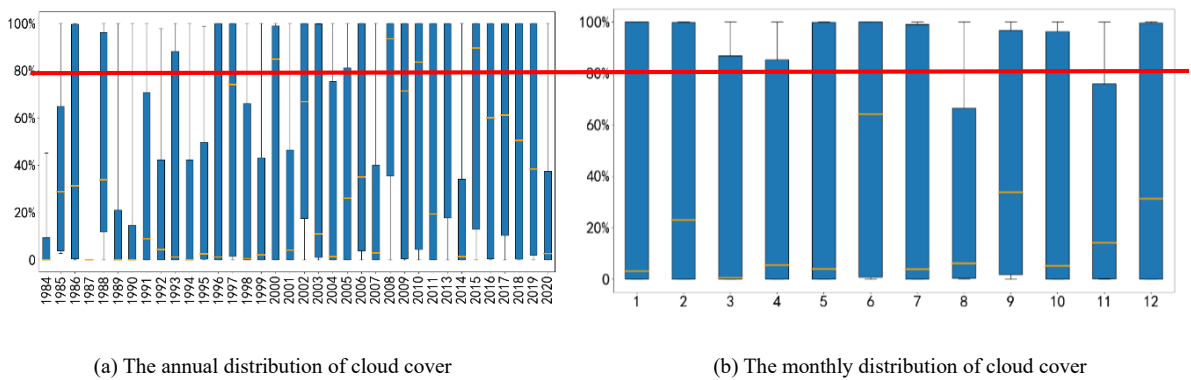

(a) The annual distribution of cloud cover  (b) The monthly distribution of cloud cover

Fig.2 Cloud covering the Miyun Reservoir area from 1984 to 2020 (images with cloud cover larger than 80% are excluded)

Landsat adopt the worldwide reference System-2 (WRS-2) path / row reference system. The spatial resolution of all bands used in this paper is 30m and the revisit period is 16 days. The band range from 0.45 μm to 2.35 μm and its settings of Landsat-5 and landsat-8 sensors are shown in Table 1, which are slightly different.

**Table 1 Comparison of Landsat-5 and Landsat-8 satellite data band settings**

| | Landsat-8 | | | Landsat-5 | |
|---|---|---|---|---|---|
| band | name | wave(μm) | band | name | wave(μm) |
| 1 | Coastal | 0.43-0.45 | | | |
| 2 | Blue | 0.45-0.51 | 1 | Blue | 0.45-0.52 |
| 3 | Green | 0.53-0.59 | 2 | Green | 0.52-0.60 |
| 4 | Red | 0.64-0.67 | 3 | Red | 0.63-0.69 |
| 5 | NIR | 0.85-0.88 | 4 | NIR | 0.76-0.90 |
| 6 | SWIR 1 | 1.57-1.65 | 5 | SWIR 1 | 1.55-1.75 |
| 7 | SWIR 2 | 2.11-2.29 | 7 | SWIR 2 | 2.08-2.35 |

**2.2.2** Digital elevation model

It can be seen from Figure 1(b) that Miyun Reservoir is surrounded by mountains, so there are a lot of mountain shadows on Landsat Image. Figure 3 shows the spectral reflectance curves of different features in Miyun Reservoir Area (based on Landsat images in autumn and winter, two hundred sample points are collected for each type of feature. The curve is the average value and the error bar is the minimum to maximum value). It can be seen that the reflectance of mountain shadow and water are low and the spectral shape has certain similarity, so mountain shadow is easily to be misclassified as water. Many studies have shown that using terrain feature can effectively remove the influence of mountain shadow (Pang kechen et al., 2016). Therefore, this paper uses the slope threshold method to effectively solve this problem. Shuttle Radar Topography Mission Digital Elevation Model (SRTM DEM) has wide coverage (60°N - 60°S) and high signal-to-noise ratio. It is widely used in the field of remote sensing detection (Wang Ling, 2000). This study uses the DEM data to calculate the regional slope of Miyun Reservoir (as shown in Figure 4) for shadow elimination near Miyun Reservoir.

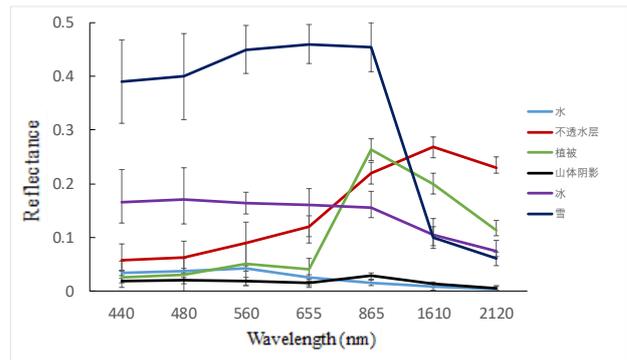

Fig.3 Spectral curves of various objects at Miyun Reservoir

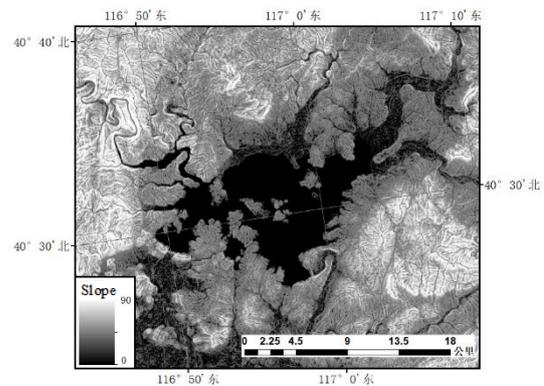

Fig.4 The slope image of Miyun Reservoir using SRTM DEM

**2.2.3 Validation data**

In this paper, two methods are used to verify the accuracy of the algorithm: 1) Direct verification, that is, select the manually interpreted samples as the test samples; 2) Cross validation, i.e. taking published water products as test samples. For each verification method, we randomly sampled a thousand sample points in Miyun Reservoir and surrounding areas. The details are as follows:

(1) Direct verification

In order to comprehensively test the water classification accuracy of the algorithm proposed in this paper under the conditions of no cloud, no ice and snow, cloud and snow cover, the images corresponding to three scenes are selected respectively: 1) no cloud, ice or snow image (October 31, 2013); 2) Cloudy image (June 24, 2011); 3) snowy and icy image (December 22, 2013). Then, based on the high-resolution Google Earth image, Landsat 8 multispectral data, MODIS global water daily mapping product (Ji et al, 2018) and other products, we judged the category of 1000 pixels and made three test samples (Table 2). The location distribution of random sample points is shown in Figure 5.

**Table 2 Numbers of the test samples used in the direct validation experiments for three conditions**

| Image | clear | | with cloud | | with Ice/snow | |
|---|---|---|---|---|---|---|
| GT | water | land | water | land | water | land |
| Nums | 394 | 606 | 386 | 614 | 401 | 599 |

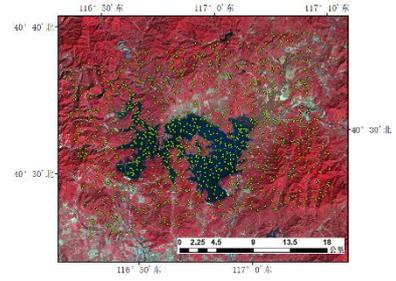

Fig.5 The distribution of randomly selected samples for the direct validation experiment.

(2) Cross validation

This study also selected the improved FROM-GLC water mapping in 2010 (Ji et al., 2015) to cross verify the results of this algorithm. Similar to direct verification, 1000 pixels were randomly sampled, including 327 water pixels and 673 non water pixels.

## 3 Research method

The purpose of this paper is to study the long-time series dynamic mapping method of Miyun Reservoir and solve the difficult problems encountered in mapping, such as cloud and ice and snow interference, terrain shadow, mixed pixels and so on, so as to realize the analysis and monitoring of the change of water surface information of Miyun Reservoir. In this paper, the Tasseled Cap Transformation is successfully applied to the cloud detection of Landsat series data. The extracted water contour and sub-pixel small targets are more accurate by using the local unmixing method and the landscape index is innovatively used to analyze the water surface morphology of Miyun Reservoir. Finally, the overall processing flow of long-time series dynamic mapping of Miyun Reservoir is proposed (as shown in Figure 6), which mainly includes the

following seven steps:

1) Cloud detection for all Landsat images;
2) Using water index to extract reservoir water;
3) Using terrain slope data to remove mountain shadow;
4) Based on the results of water extraction, snow and ice are recognized by visible band threshold segmentation;
5) The local unmixing method is used to unmix the mixed pixels at the water land boundary;
6) For cloud and snow pixels, restore the covered features through time series interpolation;
7) Based on the reservoir mapping results, the change of water surface information is analyzed, and the landscape index is calculated to analyze the shape.

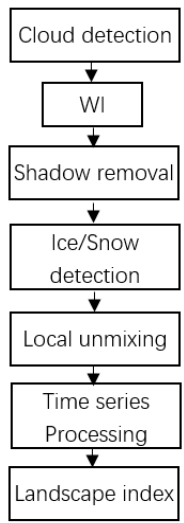

Fig.6 The flow chart of the long-term water mapping at Miyun Reservoir

### 3.1 Cloud detection

Landsat series data is often affected by cloud, resulting in the loss of surface reflectance data. The problem of cloud is the most serious in summer, as shown in Figure 2 (b). To realize the continuous mapping of the surface, it is necessary to accurately identify the cloud in the image and do time series processing.

Previous studies have shown that the fourth component of Tasseled Cap Transformation (TC) can be used for cloud detection of Landsat images (pratik et al., 2019). Tasseled Cap Transformation, also known as K-T transform (Kauth et al, 1976), is an empirical method to linearly transform the original six spectral bands (Blue, Green, Red, NIR, SWIR 1, SWIR 2) into specific six composite bands, and defines the physical meaning of the six composite bands, of which the first four bands are defined as brightness, greenness, humidity and noise respectively.

Tasseled Cap Transformation coefficient only depends on the band setting and spectral response characteristics of sensors. The data of the same sensor is fixed and not affected by the spatial and temporal aspects of the images. The calculation methods mainly include Schmidt orthogonalization and Principal Component Analysis based on statistical thinking (Wang Shuai et al. 2018). Crist(1984) proposed the Tasseled Cap Transformation coefficient of Landsat TM by using the method of principal component transformation and coordinate axis rotation. The expression of the fourth component is,

$$TC4_{TM} = -0.8242 \cdot B + 0.0849 \cdot G + 0.4392 \cdot R \\ -0.058 \cdot NIR + 0.2012 \cdot SWIR1 - 0.2768 \cdot SWIR2 \quad , (1)$$

In 2014, Baig gave the tassel cap transformation coefficient of Landsat-8 OLI data (Baig et al, 2014). The expression of the fourth component is as follows,

$$TC4_{OLI} = -0.8239 \cdot B + 0.0849 \cdot G + 0.4396 \cdot R \\ -0.058 \cdot NIR + 0.2013 \cdot SWIR1 - 0.2773 \cdot SWIR2 \quad , (2)$$

Where $B, G, R, NIR, SWIR1, SWIR2$ respectively represent the reflectivity values of Blue, green, red, near infrared, short-wave infrared 1 and short-wave infrared 2 bands. For Landsat-5, they correspond to bands 1-5 and 7, and for Landsat-8, they correspond to bands 2-7.

It can be found from formulas (1) and (2) that the coefficients of Landsat TM and OLI sensors are very similar, mainly because the central wavelength and spectral response characteristics of the two sensors are almost unanimous (see Table 1).

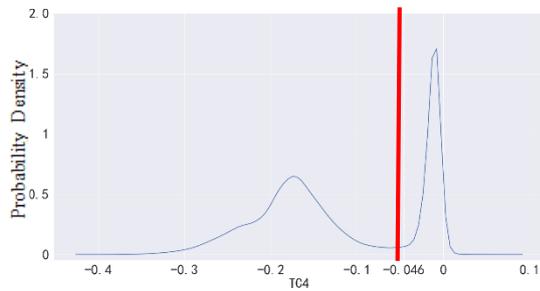

Fig.7 Distribution of Landsat-TC4 values based on 4 scenes of images at the Miyun Reservoir

Figure 7 shows the Landsat-TC4 distribution curve of Miyun Reservoir based on 4 scenes of images, including two scenes of Landsat-5 data (date: September 4, 1985, October 31, 1994) and two scenes of Landsat-8 data (date: June 1, 2018, June 27, 2018). It can be found that Landsat-TC4 data presents an obvious "bimodal distribution".

TC4 is a noise component, noise such as cloud has a large negative value on TC4, forming a peak in the negative value area, while general types of ground objects are distributed near 0, forming a steep peak near 0. In this paper, the threshold segmentation method of the fourth component of tassel transform is used for cloud recognition,

$$cloud: TC4 \leq T_{TC4} \qquad (3)$$

where $T_{TC4}$ represents the cloud detection threshold, In this paper, otsu threshold segmentation method (Otsu, 2007) is used to obtain the optimal threshold $T_{TC4} = -0.046$.

### 3.2 Water Index

As can be seen from the spectral reflectance curves of different ground objects in Miyun Reservoir area given in Figure 3: 1) the reflectance of water in visible light band (400-700nm) is relatively high, but in short wave infrared band is almost zero due to the high absorptivity of water. Therefore, the spectrum of water shows the characteristics of *high in front and low in back* in visible to short wave infrared bands;2) There is a certain fluctuation in the reflectivity value of water body in each band, that is, there is the phenomenon of *different spectrum of the same object*, which is mainly caused by the difference of water depth and water quality in different regions and periods of Miyun reservoir. In view of these two characteristics of Miyun Reservoir water body, this paper uses the water index (WI) proposed by Ji et al. (Ji et al., 2015) to extract water body, and directly compares the maximum reflectivity of short wave infrared band and visible band. The expression is as follows

$$WI = \begin{cases} 0, \max VIS \leq \max SWIR \\ 1, \max VIS \geq \max SWIR \end{cases} \qquad (4)$$

Where $\max VIS$ and $\max SWIR$ represent the maximum reflectivity in visible and short wave infrared bands respectively. For Landsat-5, $\max VIS = \max\{band1, band2, band3\}$,

$\max SWIR = \max\{band5, band7\}$. And for Landsat-8, $\max VIS = \max\{band2, band3, band4\}$, $\max SWIR = \max\{band6, band7\}$.

It can be seen from the above expression that the WI index only uses the shape characteristics of the *front high and back low* of the water spectrum and avoids the use of the spectrum value, so it can effectively avoid the impact of *the same object has different spectrum* of the water. In addition, WI is a binary index and its result can be directly used as the result of water body classification.

For images without cloud, ice, snow or other interference, WI has good extraction results and high robustness. However, like other water indexes, WI is easy to misclassify ice, snow and mountain shadow into water. Therefore, a series of subsequent processing is needed.

**3.3 Remove mountain shadow**

Mountain shadow is very easy to be misclassified as water. It can be seen from the spectral curve in figure 3 that the spectral curve of mountain shadow and water is very similar, and the overall reflectivity of both is low. It is found that nearly 80% of terrain shadows are misclassified as water by WI. Therefore, reservoir mapping needs to further eliminate the influence of mountain shadows.

In China, wetlands (including mainly water) with slopes less than 3 ° and 8 ° account for 93.85% and 99.17% of the total wetland area respectively (Niu Zhenguo et al., 2009). Figure 8 shows the mapping results of Miyun Reservoir according to FROM-GLC in 2017(http://data.ess.tsinghua.edu.cn/fromglc10_2017v01.html) statistical slope distribution curve. It can be seen that almost all the water bodies of Miyun Reservoir are distributed in the flat area below the slope of 4°. Therefore, this paper uses the slope threshold method to eliminate the mountain shadow in the water body results, that is

$$shadow: slope > T_{slope} \qquad (5)$$

Where $T_{slope}$ indicates the slope threshold. Pixels with a slope greater than 4° are excluded from the results of water.

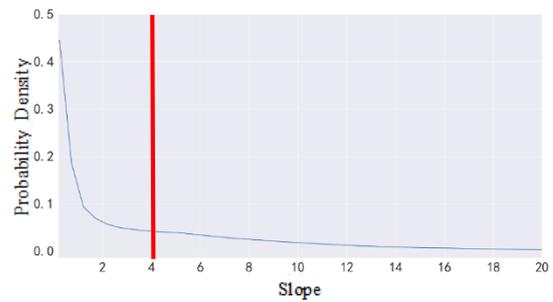

Fig.8 Distribution of slope based on 2017 FROM-GLC statistics at Miyun Reservoir

**3.4 Snow and ice detection**

According to the spectral reflectance curves of different ground objects in Miyun Reservoir area given in Figure 3, the snow and ice spectrum also has the spectral characteristic of *high in front and low in back*, therefore, the WI index will mistakenly classify it as water. However, it can also be seen that the reflectance of snow and ice in the visible band is much higher than that of water. Taking advantage of this feature, this study uses the reflectance threshold segmentation method in the visible band to extract the wrong snow

and ice pixels in the water results. Specifically, if a pixel satisfies the following condition:

$$ice/snow: \max VIS \geq T_{\max VIS}. \quad (6)$$

Then mark it as ice and snow, where the threshold is set as $T_{\max VIS} = 0.15$.

**3.5 Local unmixing**

Affected by the resolution of Landsat sensor, there are a large number of mixed pixels at the water land boundary. These mixed pixels affect the accurate calculation of reservoir area, interfere with the accurate extraction of water land boundary, and also affect the research on the landscape form of reservoir boundary.

Usually, the mixed pixel problem is solved by linear spectral unmixing method. For the mixed pixel of water and land, it can be linearly expressed by the water and land dual endmember model, and the endmember coefficient meets the constraints of nonnegative and sum of 1,

$$\mathbf{r} = c_W \mathbf{e}_W + c_L \mathbf{e}_L \quad (7)$$

Constraint one: $c_W + c_L = 1 \quad (8)$

Constraint two: $c_W, c_L \geq 0 \quad (9)$

Where, $\mathbf{r}$ represents the reflectance vector of mixed pixels, $\mathbf{e}_W, \mathbf{e}_L$ represent the reflectance vectors of water and land end members respectively, and $c_W, c_L$ represent the abundance coefficients of end members in a single mixed pixel.

If the water and land endmember corresponding to each mixed pixel are known, the water and land abundance in the mixed pixel can be calculated by spectral unmixing. Therefore, how to determine the water and land endmember for each mixed pixel becomes the key. Spectral unmixing analysis usually uses the pure pixel endmember of a variety of ground objects to linearly decompose the mixed pixels (Nascimento et al, 2005; Geng et al, 2013; Ji et al, 2015). However, in practice, there are many types of satellite data and ground objects, the pure endmembers are difficult to obtain due to the influence of sensor resolution. Taking Miyun Reservoir as an example, the pure water endmember on both sides of the water land boundary may be shoal water pixels or deep-water pixels. There are many types of land endmember, such as bare land, farmland, grassland, forest and impervious layer. It is very difficult to obtain these endmembers spectra using Landsat 30m resolution data. In order to solve the problem, we use the local unmixing method (Ji et al., 2015).

The specific steps of using the local unmixing method to calculate the water abundance at the boundary are as follows: 1) determine the water land mixing area. Here, the 8 neighborhood pixels connected to the land and water boundary pixels are defined as the mixing region (green pixels in Fig. 9);

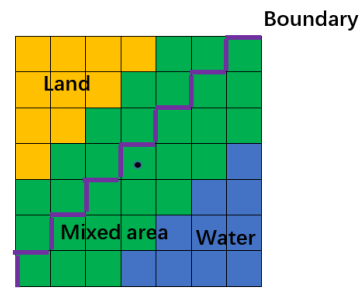

Fig.9 Schematic diagram of the mixed area

2) Search the water and land endmembers. We search the water and land endmembers in a 5 × 5 window centered on the mixed image element. To ensure that the water and land endmembers are pure pixels as much as possible, we select the pixels with the lowest and highest reflectance in the window as the water and land endmembers; 3) The abundance of water for mixed image elements can be obtained by calculating the least squares decomposition under full constraints (Geng et al, 2016); 4) Threshold segmentation. The mixed image element satisfying the threshold is classified as a water pixel, otherwise it is a non-water pixel.

### 3.6 Time Series Processing

Most of the previous mapping studies treat cloud, snow and ice more coarsely, for example, the global month-by-month water product JRC-water released by Pekel in 2016 (Pekel et al, 2016) treated clouds, snow and ice as no remotely sensed data, resulting in a large number of missing results in its mapping in cloudy season and winter. In this paper, to exclude the interference caused by clouds and snow and ice, a time series interpolation method is used for processing. First, for each spatial point in the preliminary mapping results, a time series vector is established, $\mathbf{X} = [x_1, x_2, ..., x_N]$, N is the number of image time phases, and $x_i$ represents the classification results of the pixel points in time phases $i$, there are four cases as follows.

$$x_i = \begin{cases} 1, water \\ 0, land \\ C, cloud \\ I, ice/snow \end{cases} \quad (10)$$

In this paper, we consider the first two (1 and 0) as valid data and the last two (C and I) as invalid data. For any moment of $x_i$, if $x_i = C$ or $x_i = I$, we search its before and after time neighborhood classification results and interpolate them with the valid classification results (1 or 0) with the nearest time interval.

### 3.7 Landscape Index

Landscape index can be used to describe landscape patterns and establish links between landscape structures and processes or phenomena to better explain and understand landscape functions. In recent years, the study of wetland landscape patterns has become a hot issue in the field of landscape ecology and global change research, and describing landscape patterns through landscape indices and comparing and analyzing the characteristics of patterns at different scales have long been of great interest to landscape ecologists (Fu, 1995; Yang Yongxing, 2002; Song Minmin et al., 2018). Miyun Reservoir is divided into two parts, East and West (Liang Xiujuan et al., 2005; Li Huimin et al., 2007). Over the years the reservoir morphology has been influenced by the amount of water, with the East and West reservoirs constantly splitting and unifying. In this paper, we use landscape division index to describe the fragmentation of the reservoir waterscape and try to capture the changes between the east and west reservoirs. Landscape division degree is an index that describes the probability that two randomly selected pixels in the landscape are not in the same patch (Li et al., 2019). In this study, the 8-connected algorithm is used to calculate landscape division degree $P_D$, and the formula is as follows:

$$P_D = 1 - \sum_{k=1}^{M} \frac{S_k^2}{S^2}, \quad M \geq 1, \quad (11)$$

$S$ is the total area of the landscape, $S_k$ is the area of the kth disconnected patch in the landscape, $M$ is the total number of patches. Figure 10 shows the corresponding $P_D$ examples of several different landscape forms. It can be seen that when the landscape forms as a whole, $P_D = 0$; When landscape form M equally divided, $P_D = 1 - 1/M$. Therefore, the larger $M$ is, the larger $P_D$ is, that is, the more pieces the landscape is divided into, the larger the landscape division index is; When the landscape is divided unequally into M patches, $0 < P_D < 1 - 1/M$ （$M \geq 2$）。

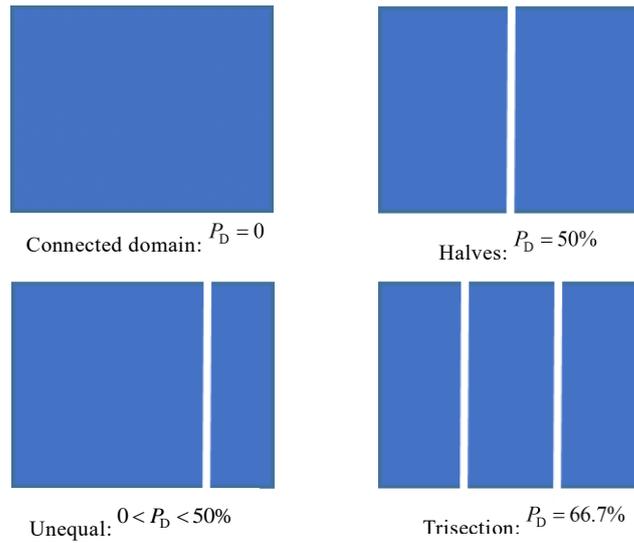

Connected domain: $P_D = 0$
Halves: $P_D = 50\%$
Unequal: $0 < P_D < 50\%$
Trisection: $P_D = 66.7\%$

Fig.10 Examples of Division of different landscape shapes

### 3.8 Precision

In order to quantitatively evaluate the reservoir mapping results, we calculated the Overall Accuracy (OA), Precision and Recall, as shown in Formula (12) - (14). All three can be calculated according to the confusion matrix (Table 3).

**Table 3 Confusion matrix for two categories**

|  | True | False |
|---|---|---|
| Prediction(T) | TP | FP |
| Prediction(F) | FN | TN |

$$OA = \frac{TP + TN}{TP + TN + FP + FN} \quad (12)$$

$$precision = \frac{TP}{TP + FP} \quad (13)$$

$$recall = \frac{TP}{TP + FN} \quad (14)$$

## 4 Results

### 4.1 Reservoir mapping results

The reservoir mapping results obtained based on the algorithm of this paper are shown in Figure 11. By comparison with Figure 11 (a, b), it can be seen that water of the reservoir is well extracted. According to Figure 11 (c, d), the water and the mid-lake island are well distinguished, the details of the water boundary are well retained, and the shadow of the mountain is also eliminated. Figure 11 (e, f) for reservoir in northern area, you can see the long and narrow cofferdam (marked by arrow) in the reservoir is extracted, the actual maximum width does not exceed 25m (by visual interpretation based on Google Earth Pro high-resolution image), which is smaller than the image resolution (30m), actually the local unmixing is helpful for extraction.

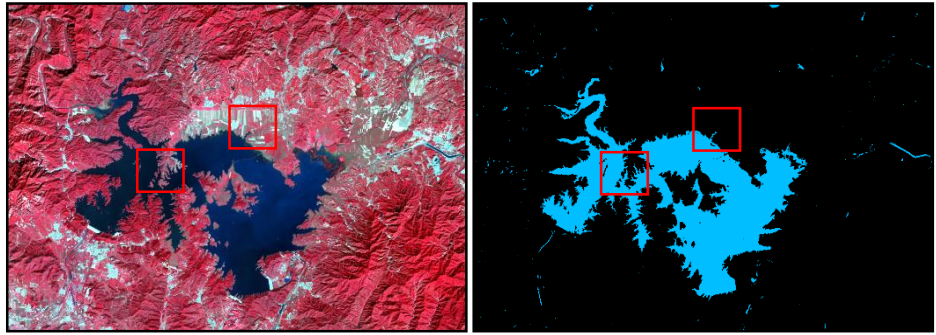

(a) False color image of Miyun Reservoir    (b) Mapping result

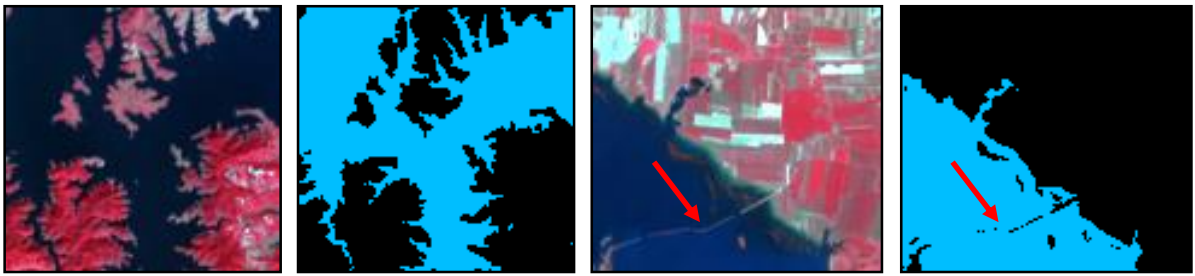

(c) False color image of center islands    (d) Zoom of center islands    (e) False color image of northern area    (f) Zoom of northern area

Fig.11 Mapping results of Miyun Reservoir (date: 2013.10.03)

For images with clouds, the water index (WI) will misclassify the clouds, and the WI extraction results are shown in the red rectangular box in Figure 12(b). The tasselled cap transformation and threshold segmentation used in this paper can effectively identify clouds (Figure 12(c)), and then after time series interpolation processing, the water under the clouds can be recovered (Figure 12(d)). But due to the error of cloud detection method, the undetected clouds are classified as ground (shown in the dashed box area in Figure 12(d)), which leads to the decrease of the accuracy of mapping. The extraction accuracy of the images with clouds is detailed in section 4.2.

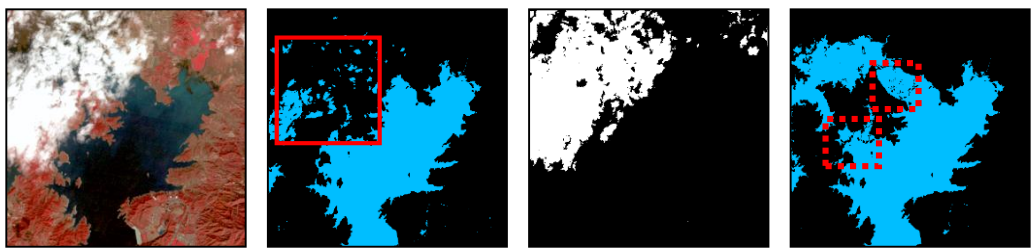

(a) False color image with cloud    (b) WI result    (c) Cloud extracted in this paper    (d) Water map after time series processing

Fig.12 The recovered result of cloud image after time series processing (date: 2010.10.11)

## 4.2 Accuracy verification

### 4.2.1 Direct verification

The accuracy of the algorithm in this paper for three cases (imge without cloud or snow/ice, image with clouds, image with snow/ice) is shown in Table 4, and it can be seen that, 1）Imge without cloud or snow/ice. The overall accuracy is 98.2%, especially the accuracy rate reaches 98.4%, indicating that the algorithm in this paper can extract the water of Miyun Reservoir accurately by using image without clouds and ice.

Table 4 The accuracy for direct verification

| Image | OA(%) | Precision(%) | Recall(%) |
|---|---|---|---|
| cloud | 98.2 | 98.4 | 97.0 |
| With cloud | 91.4 | 89.8 | 88.0 |
| With ice/snow | 95.8 | 92.8 | 97.0 |

2）Image with clouds. Among the three cases, cloud has the greatest impact on the accuracy of water extraction, and the accuracy and recall rate are lower than 90%, which is mainly due to the error of cloud identification method, which makes some water pixels not extracted (e.g., Figure 12(d)), true positive (TP) samples decreased, resulting in the reduction of various accuracy indicators.

3）Image with snow/ice. The recall rate is highly reaching 97%, but the accuracy rate is relatively low, at 92.8%. The reduced classification accuracy of the images is due to the fact that a part of snow and ice are classified as water in the threshold segmentation process between snow/ice and water. But part of the snow/ice may actually be on land, resulting in the increase of false positive (FP) samples, which leads to a decrease in accuracy.

### 4.2.2 FROM-GLC cross-validation

Table 5 shows the confusion matrix obtained by cross-validation using the improved FROM-GLC water layers, in which overall extraction results were 99.4% the same. The results show that the consistency between this paper's water body extraction results and FROM-GLC is very high, which actually shows that the proposed method can be comparable to the accuracy of water extraction methods based on manual selected training samples and supervised classification (support vector machine). The method in this paper does not need to manually select samples for each scene image, so the method in this paper is more suitable for high-precision reservoir dynamic mapping of long time series.

Table 5 The accuracy for cross validation

| | | Mapping | | |
|---|---|---|---|---|
| | | W | L | Precision(%) |
| FROM-GLC | W | 321 | 6 | 98.2 |
| | L | 0 | 673 | |
| | Recall(%) | 100 | 99.4 | |

Notice: W represents water, L represents land.

# 5  Analysis and discussion on water area information of Miyun Reservoir

## 5.1  Interannual variation in Miyun Reservoir

Figure 13 gives a map of the water coverage rate of Miyun Reservoir from 1984 to 2020, which can be seen intuitively that the water area of Miyun Reservoir has undergone great changes, among which the areas with greatest changes are the northern area, the inlet of the Chao River and the Bai River, and the mid-lake island. During the wet period, the reservoir expands to the north and the inlet of the Chao Bai River is covered by water. The mid-lake island is mostly submerged and only a small part of it is left. While during the dry period, the reservoir area shrinks severely and the mid-lake island is exposed.

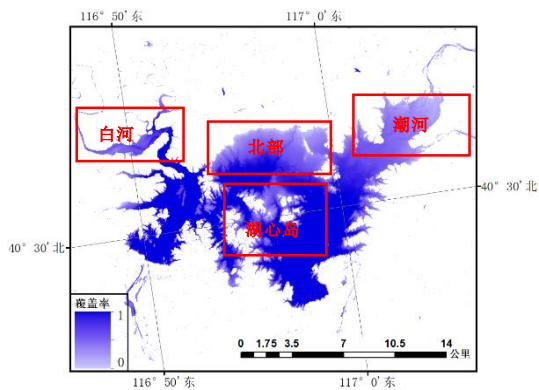

Fig.13 Water coverage of Miyun Reservoir (1984-2020)

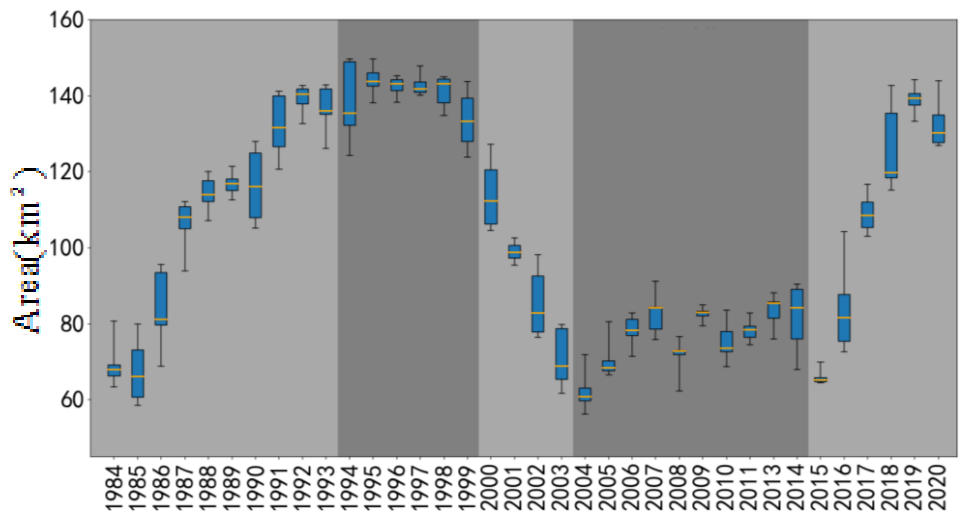

Fig.14 Interannual change of the area of Miyun Reservoir (1984-2020)

Figure 14 shows the changes of Miyun Reservoir area (the ares of inlet and outlet of the Chao river and the Bai river are not counted, the same goes for the following) from 1984-2020 (data in 2012 is missing due to the absence of satellite service) The box properties in the figure are upper boundary, upper quartile, median, lower quartile, and lower boundary. The water area of Miyun Reservoir has changed greatly in the past 37 years. Its maximum area is 151.6 km$^2$ (1995), minimum area is 57.3 km$^2$ (2004) and recovering to 148.0 km$^2$ in 2019 (2019.09.18), which is consistent with the Beijing Bureau of Landscape Architecture (http://yllhj.beijing.gov.cn/sdlh/ jjjlylhxtfz/201909/t20190929_531708.shtml) published monitoring results(maximum area is 161.1 km$^2$, minimum area is 56.4 Square kilometers and recovered to 148.7 km$^2$ in 2019 (2019.08.15)). Based on the changes in the water surface area of Miyun Reservoir, it can be divided into five periods: the "growth period" (1984-1993), the "peak period" (1994-1999), the "decline period" (2000-2003), the "protection period" (2004-2014), and the "recovery period" (2015-2020). Specifically,

The first period: the "growth period" (1984-1993): the reservoir area grew rapidly due to a series of policies promulgated by the government to control and protect the over-consumed Miyun Reservoir. In 1985, Beijing issued the regulations of "Two storages and one Canal" (Zhu Zuxi, 1986), which prohibited the construction of any projects except water conservancy projects. A large number of enterprises were closed down and the projects under construction were suspended. Under the influence of the policy, the ecological environment of Miyun Reservoir has been improved and the water area has been expanded.

The second period: the "peak period" (1994-1999): the reservoir area maintained at the peak level. While the government continued to implement the reservoir protection policy, it also took measures to improve the reservoir ecological environment through reservoir migration. By the early 1990s, the population in the area around the reservoir was seriously overloaded. It took 6 years to emigrate a total of 12,484 people from the reservoir area, and the problem of population overload in the surrounding area of Miyun Reservoir has been greatly alleviated. During this period, the ecological environment and water quality of Miyun Reservoir kept improving. Thanks to the national protection policy, the reservoir area reached the peak of 151.6 km$^2$ around 1995 and remained at the peak level during this period.

The third period: the "decline period" (2000-2003): During this period, the reservoir area showed a rapid decline, mainly influenced by arid climate and urban expansion water consumption. At that time, Beijing was experiencing consecutive years of drought and Figure 15 shows the precipitation curve of Beijing from 2000-2018 drawn by using data from the National Earth System Science Data Center (Peng et al, 2019). It can be seen from the figure that the annual precipitation in Beijing during the decline period (424.48 mm) was much lower than the average annual precipitation from 2000-2018 (488.42 mm). In fact, according to the observation of 20 meteorological stations in Beijing, the average annual precipitation in the five years from 1999 to 2003 is the lowest value since 1961, only 442.7mm. 141.9mm less than the average annual precipitation since 1950

(Xu Zongxue et al., 2006; Li Yongkun et al., 2013), the average annual precipitation of the whole city is only 2/3 of the annual average precipitation, the surface water supply accounts for about 1/3 of the annual average surface water supply, and the groundwater supply in plain area accounts for about 1/2 of the annual average groundwater supply, with an annual water supply and demand gap of about 1.5 billion m$^3$ (Wang, 2004). Therefore, the water resources situation in Beijing is very severe. The drought and large water shortage have put the Miyun Reservoir in crisis, which covers only 57.3 km$^2$ at minimum.

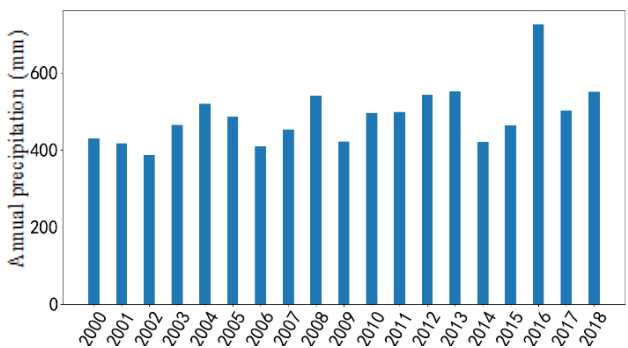

Fig.15 Annual precipitation in Beijing (2000-2018)

The fourth period: the "protection period" (2004-2014): during the "protection period", the area of Miyun Reservoir remained low and stable, mainly due to the influence of the national protection policies again. In 2004, Wang Jinru presented an analysis of the current situation and outlook of Beijing's water resources at the 8th Cross-Strait Water Resources Science and Technology Exchange Conference (Wang Jinru, 2004). In order to solve the water supply problems of Beijing Olympic Games in 2008, the state once again implemented a protection policy for Miyun Reservoir, then the Miyun Reservoir area of the declining trend under control, the area of the reservoir remained at a relatively stable level over a ten-year period. (Jia Dongmin et al., 2012).

The fifth period: the "recovery period" (2015-2020): the reservoir area gradually recovered and recently approached the peak level; At 14:32 pm on December 12, 2014, the South-North Water Diversion Project was officially opened to the public and the reservoir area continued to climb due to the influence of river water entering the reservoir. In addition, it can be seen from Figure 15 that precipitation in Beijing also increased during the recovery period (precipitation in 2016 exceeded 700 mm, and average precipitation in other years (505.58mm) was slightly higher than the average precipitation in 2000-2018 (488.42mm) except 2016). so it also contributed to the recovery of reservoir water. Therefore, during the recovery period, Miyun Reservoir was mainly affected by the South-North Water Diversion Project. At the same time, due to the increase of rainfall, short-term rainstorm and the protection policy of reducing the outflow of water, the reservoir area has recovered to nearly the peak value in 1995.

In summary, the long-term changes of Miyun Reservoir area were mainly influenced by policies, water conservancy projects, climate, urban construction, and industrial water use and other factors. In the five periods during the 37 years of Miyun Resevoir. The main reasons for the decline period are continuous drought and excessive water resources consumed by urban construction. The recovery period is mainly affected by the South-North Water Diversion Project, as well as by protection policies and continuous heavy rains. Therefore, policy, water conservancy project and climate are the main driving factors affecting the change of Miyun Reservoir area.

**5.2** Annual changes of Miyun Reservoir

In order to study the change of water area change in Miyun Reservoir. According to Figure 14, we selected 1985 (growth period), 1995 (peak period, with the largest area), 2002 (decline period), 2004 (protection period, with the smallest area), and 2017 (recovery period) to make water cover maps in each year, as shown in Figure 16. It can be seen that when the reservoir is full of water in a year, it will expand to the area with lower elevation in the north, and the area of inlet of the Chao river will also increase significantly. In 1985 (Figure 16(a)) and 2002 (Figure 16(c)), the mid-lake island area changed greatly. In 2017 (Figure 16(e)), obvious changes also occurred in the southern region of the reservoir. Specifically：

- In 1985 (Fig. 16(a)), due to the lack of water in the reservoir, Miyun Reservoir was divided into two regions: east and west. However, in September, when the water was the most abundant, the east and west of the reservoir merged into one. In fact, there is a man-made cofferdam between the east and west (Figure 11 (e, f)). When the reservoir water level is low, the east and west can exchange water through the gap in the cofferdam (Li Huimin et al., 2007).

- In 1995 (Figure 16(b)), the water volume of the reservoir was sufficient and there were obvious area changes in the inlet of the Chao river and the northern area.

- In 2002 (Figure 16(c)), except for the northern area and the inlet of the Chao river, there was also an area change of the mid-lake island. Since it was in the decline period at that time, the reservoir area has been shrinking year after year, so the area of the mid-lake island has been expanding, resulting in the change of water coverage around the mid-lake island.

- In 2004 (Figure 16(d)), the reservoir was at the point in time when the area was the smallest, the eastern and western parts was clearly distinguished. In addition to the obvious changes in the inlet of the Chao river, between the eastern and western parts also through a long and narrow diversion channel for water exchange. Because the reservoir area was too small and the water level was too low to reach the height of the cofferdam, the two parts can only be connected through the diversion canal.

- In 2017 (Figure 16(e)), in addition to the northern region and the inlet of the Chao River, due to the increase of water volume in the reservoir during the recovery period, the Westone Camel Sub-dam area in the south also had been an obvious water expansion.

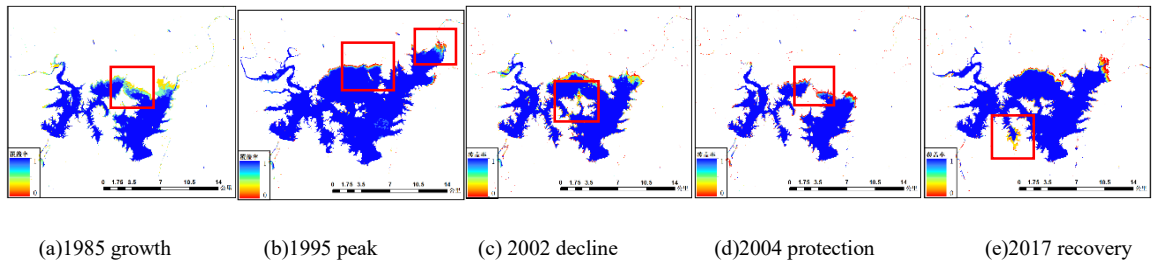

(a)1985 growth   (b)1995 peak   (c) 2002 decline   (d)2004 protection   (e)2017 recovery

Fig.16 (a), (b), (c), (d), (e) are the coverage maps in 1985, 1995, 2002, 2004, and 2017

Further, we counted the months when Miyun Reservoir reached the maximum and minimum area each year. From Table 6, we can see the area of Miyun Reservoir reached the maximum in August and September much more often than in other months, while the minimum in May much more often than in other months, so we can assume that the area of Miyun Reservoir reached the maximum in August-September and the minimum in May. This is generally consistent with the findings of Ji et al. on the analysis of global water body area change, which pointed out that the maximum area of global water bodies appeared in September (Ji et al., 2018). However, this article also indicated that the minimum area of water appeared in February, while the result of our research is in May, and the main reason for the large difference is that Ji considered snow and ice in winter as non-water and did not include it when counting the area of water, while in this paper, the time series of the frozen area of Miyun Reservoir in winter is also counted after processing.

Therefore, the minimum area does not occur in winter but in May. Obviously, this result is more consistent with the Miyun Reservoir. Since Miyun Reservoir is located in the warm temperate continental monsoon climate zone, the rainy season is from July to August and its precipitation accounts for 70% of the year (Li Miaomiao et al., 2004). Therefore, when the rainy season ends in August and September, the area of Miyun Reservoir accumulates a large amount of rain, which reaches the largest area in a year. After the end of the rainy season each year, as rainfall decreases and the upstream river is dry all year round (Li Huimin et al., 2007), the downstream of the reservoir continues to release water at the different time (Zhang, Jiantao et al., 2010), the area of Miyun Reservoir continues to decline. In addition, before the coming of the rainy season, miyun Reservoir will lower the water level to play the flood control function in the rainy season. Therefore, the reservoir area will reach its minimum in May each year.

**Table 7 Number of times that the water area reaches the maximum and minimum for the Miyun Reservoir**

| month | 1 | 2 | 3 | 4 | 5 | 6 | 7 | 8 | 9 | 10 | 11 | 12 |
|---|---|---|---|---|---|---|---|---|---|---|---|---|
| max(nums) | 1 | 0 | 0 | 0 | 0 | 0 | 1 | 13 | **19** | 1 | 0 | 1 |
| min(nums) | 1 | 1 | 0 | 3 | **28** | 1 | 0 | 2 | 0 | 0 | 0 | 0 |

### 5.3 Landscape morphological changes of Miyun Reservoir

Figure 17 shows the interannual variation trend of landscape separation index of Miyun Reservoir. It can be seen that the landscape separation index $P_D$ has been distributed around 0 or 50% for many years, but its value is slightly higher. This is because in the year when the reservoir is short of water (such as 2004), the area shrinks and the reservoir breaks off from the mid-lake island area and divides into two reservoir areas, east and west (Figure 16(d)). According to the definition of landscape separation index, the index is close to 50%, indicating that the reservoir areas in the east and west of the reservoir are equivalent. However, in the years when the reservoir has a large amount of water (such as 1995), the east and west of the reservoir are connected as a whole, and the landspace division index is small and the is close to 0. It can be seen from Figure 17 that the east and west reservoir areas of Miyun Reservoir experienced a process of *division-combination-division-combination* in 37 years. Combined with Figure 14, it can be seen that the reservoir was in the form of *combination* in most periods, and the *division* mainly appeared in the early protection period (1984 and 1985) and the decline period (2004). Compared with the change of reservoir area, the change of landscape fragmentation index can further monitor the separation and integration state between the eastern and western parts and provide additional morphological information such as the area ratio of the eastern and western parts. Landscape morphology is often the macro reflection of the reservoir water quantity, ecological environment and other factors. Therefore, landscape division index can be used as one of the important indicators for resource information monitoring and management of Miyun Reservoir.

Landscape index can truly reflect the macro change process of reservoir area. The introduction of landscape index provides a new scientific perspective for water resources monitoring and management in the future. It is hoped that more types of landscape index will be applied to water resources scientific monitoring in the future.

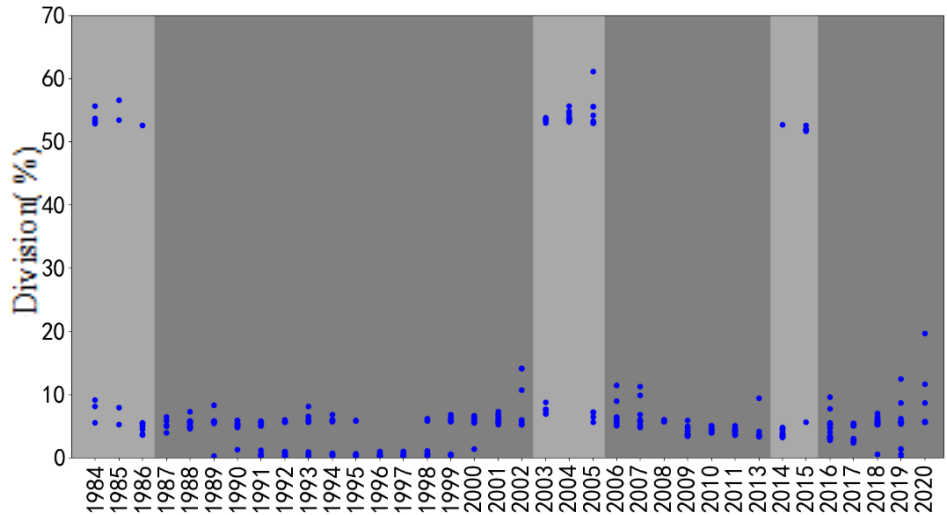

Fig.17 The interannual change of the landscape division index (1984-2020)

# 6 Conclusion

In this paper, based on the Landsat data to Miyun Reservoir as the study area, we studied the difficulties in the long time series dynamic water mapping, such as cloud, snow/ice interference, terrain shadow and mixed pixel, and successfully applied the Tasselled Cap Transformation to Landsat reflectivity data for cloud detection, and used local unmixing method to effectively solve the mixed image element problem in large-scale mapping, and innovatively used landscape to analyze the changes of water area morphology. Finally, we propose a complete and efficient reservoir mapping process, including: 1) cloud detection, 2) water extraction, 3) shadow removal, 4) ice and snow detection, 5) local unmixing, and 6) time series interpolation. Using this algorithm to complete dynamic mapping of Miyun Reservoir from 1984 to 2020. The accuracy of the mapping results is high and the overall accuracy of direct verification is 98.2% in the case of no cloud and no snow/ice, and the overall consistency with the existing water mapping products (improved FROM-GLC) cross verification is up to 99.4%. Based on the long-term series analysis of water surface information such as area, coverage rate and morphological characteristics of Miyun Reservoir, the following conclusions are drawn:

1) The area of Miyun Reservoir changed greatly during the 37 years from 1984 to 2020, with a maximum of 151.6 km2 and a minimum of 57.3 km2. The area changes mainly concentrated in three regions, including the northern area, the inlet of the Chao river and the Bai river, and the mid-lake island;

2）Based on the area change of Miyun Reservoir, it can be divided into five periods: the "growth period" (1984-1993), the "peak period" (1994-1999), the "decline period" (2000-2003), the "protection period" (2004-2014), and the "recovery period" (2015-2020);

3) The annual variation of Miyun Reservoir mainly occurs in four regions, including the northern area, the inflow area of the Chao river, the mid-lake island, and the Westone Camel Sub-dam. The annual maximum area occurs in August and September, and the minimum area occurs in May, because the reservoir drains in May before the annual rainy season, resulting in the minimum area. At the end of the rainy season in August and September, a large amount of rainwater is accumulated, and the area reaches the largest in a year;

4) The landspace division index of Miyun Reservoir changed greatly in 37 years, and it was divided into the east and west when the water volume was small. From 1984 to 2020, the eastern and western parts experienced the process of *division-combination-division-combination*, including 1984-1986, 2003-2005 and 2014-2015, and the other years were in the merger stage.

In the following research, we will continue to pay attention to Miyun Reservoir by applying the reservoir mapping method in this paper. Try to apply it to other reservoirs and monitoring the changes of water surface information of reservoirs in China on a larger scale.

## 参考文献(References)